\documentstyle[aps,prb]{revtex}
\tighten

\begin{document}

\twocolumn[\hsize\textwidth\columnwidth\hsize\csname @twocolumnfalse\endcsname

\title{ Majorana fermion formulation of the two channel Kondo model}
\author{G.-M. Zhang, A.C. Hewson and  R. Bulla}
\address{Department of Mathematics, Imperial College, London SW7 2BZ, UK.}
\maketitle

\begin{abstract}
{We show that a Majorana fermion description
of the two channel Kondo model can emerge quite naturally 
as a representation of the algebra associated with the spin currents
in the two channels. Using this representation we  derive an exact equivalent 
Hamiltonian
for the two channel model expressed entirely in terms of 
Majorana fermions. The part of the Hamiltonian that is coupled 
to the impurity spin corresponds to the vector part of 
the $\sigma$-$\tau$ model (compactified two channel model).
Consequently all the thermodynamic properties associated with the impurity
spin can be calculated from the $\sigma$-$\tau$ model alone.
The equivalent model can be used to confirm the interpretation of the
many-body excitation spectrum of the low energy
fixed point of the two-channel model as due to free Majorana fermions with 
appropriate boundary conditions.}
\end{abstract}
]


The two channel Kondo model is known  to have
a low energy non-Fermi liquid fixed point and
has been put forward as a model to explain non-Fermi liquid behavior 
as observed in several quite different physical systems at low temperatures, 
such as certain heavy fermion alloys and two-level systems. A full description 
of the model and the various theoretical approaches that have been applied 
to  elucidate its physics, plus
its potential applications, can be found in a recent extensive and thorough
review by Cox and Zawadowski \cite{cz}. There are exact
solutions for the ground state and thermodynamics of the model
from Bethe ansatz calculations \cite{andrei} which have been known for some 
time, but there are continuing  efforts to find a simple intuitive 
understanding of the 
nature of the excitations in the neighborhood of the low energy fixed point.
Numerical renormalization group \cite{pc} and conformal field theory  
calculations \cite{al}
 give predictions for the many-body excitations at the fixed point
but they do not provide a simple explanation of these excitations in terms 
of more elementary ones,
as is possible at the Fermi liquid fixed point of the single channel Kondo
model. 

The bosonization approach of Emery and Kivelson \cite{ek} 
showed that close to a particular value of the coupling in the strong coupling 
regime (analogous to the Toulouse limit of the one channel model),
 the low temperature behavior has  the same form as the weak 
coupling model at low temperatures with the same Wilson ratio.
This suggested that the low energy fixed point could be described
by some effective Hamiltonian of this form with renormalized parameters.
At the solvable Emery-Kivelson point the effective Hamiltonian contains a 
combination of particle and hole creation operators which can be expressed 
most conveniently in terms of Majorana fermions. Recently, 
Maldacena and Ludwig using abelian bosonization have reformulated 
the conformal field theory in terms of Majorana fermions \cite{ml}. 

The Majorana fermion approach was developed further by
Coleman, Ioffe and Tsvelik
\cite{cit} who introduced the `compactified' two channel 
or $\sigma$-$\tau$ model. This is a single channel model in which the impurity 
spin is coupled to  the conduction electron spin and isospin, a combination 
which is most conveniently expressed in terms of three Majorana fermions.
It was conjectured, with supporting arguments, that the 
$\sigma$-$\tau$ model has the same low energy fixed point as the two channel 
Kondo model on which it was modeled. Subsequently Coleman and Schofield 
\cite{cs} introduced a version of the Anderson model (the O(3) or compactified 
Anderson model \cite{zh}) which can be mapped into the $\sigma$-$\tau$ model 
in the local moment regime using the Schrieffer-Wolff transformation. 
Recent numerical renormalization 
group, conformal field theory, weak and renormalized perturbation 
expansion results for both the $\sigma$-$\tau$ and O(3) Anderson models of 
Bulla, Hewson and Zhang
\cite{bh,bhz} confirm that the localized compactified models do give the same 
low temperature thermodynamic behavior as the two channel model.

In this paper we show that the 
Majorana fermion description can emerge quite naturally within the two channel 
model as a representation of the algebra of the total spin current of the two 
channels. We further show using this representation that an exact equivalent 
model for the two channel Kondo model can be obtained  entirely in terms
of Majorana fermions. The part of this model which includes the interaction
with the impurity spin is the vector part of the $\sigma$-$\tau$ model.
This implies that the impurity spin contribution to the thermodynamic 
properties of 
the two channel Kondo model can be calculated from the $\sigma$-$\tau$ model
alone. We show that this Majorana fermion version of the model can 
 be used to confirm the interpretation of the low energy fixed point of the 
two channel model as free Majorana fermion excitations subject to appropriate 
boundary conditions.


We start with the Hamiltonian in the form,
\begin{eqnarray}
 && H= H_0+H_I  \nonumber \\ 
 && H_0=\frac{v_f}{2\pi}\sum_{j=1}^{2}\sum_{\sigma=\uparrow,\downarrow}
    \int_{-\infty}^{+\infty}dx 
    :\psi^{\dag}_{j,\sigma}(x)(i\partial_x)\psi_{j,\sigma}(x): 
\nonumber \\ 
&& H_I=\sum_{a=x,y,z}J_a S_d^a J_s^a(0),
\end{eqnarray}
where we have retained the s-wave scattering only, linearized the fermion 
spectrum, and replaced the incoming and outgoing waves with two 
left-moving electron fields $\psi_{j,\sigma}(x)$; $J_s^a(x)$ are the 
conduction electron spin current operators
\begin{equation}
   J_s^a(x)= \sum_{j,\sigma,\sigma'}
     :\psi^{\dag}_{j,\sigma}(x)s^a_{\sigma,\sigma'}\psi_{j,\sigma'}(x):  
\end{equation}
$s^a$ being spin-1/2 matrices. We can also introduce charge and flavor
currents
\begin{eqnarray}
&& J_c(x)=\sum_{j,\sigma}:\psi^{\dag}_{j,\sigma}(x)\psi_{j,\sigma}(x): 
\nonumber \\ 
&& J_f^a(x)=\sum_{j,j',\sigma}:\psi^{\dag}_{j,\sigma}(x)t_{j,j'}^a 
                     \psi_{j',\sigma}(x):.  
\end{eqnarray}
where $t_{j,j'}^a$ are generators of an SU(2) symmetry group. Following Affleck
and Ludwig \cite{al}, the free part of the Hamiltonian can be rewritten as a 
sum of three commuting terms by the usual point-splitting procedure
(Sugawara construction): 
\begin{eqnarray}\label{suga}
 H_0=\frac{v_f}{2\pi}\int_{-\infty}^{+\infty}dx
    && \left [\frac{1}{8}:J_c(x)J_c(x):+
     \frac{1}{4}:\vec{J}_f(x)\cdot\vec{J}_f(x):\right.
\nonumber \\       
     && \left .+\frac{1}{4}:\vec{J}_s(x)\cdot\vec{J}_s(x): \right ],
\end{eqnarray}
while the interaction term is expressed in terms of the electron spin 
currents and the impurity spin only. The information about the number of 
channels is contained in the commutation relations obeyed by the spin currents
\begin{equation}
 [J_s^a(x), J_s^b(x')]
  =i\epsilon^{abc}J_s^a(x)\delta(x-x')+\frac{ki}{4\pi}\delta_{a,b}\delta'(x-x')
\end{equation}
indicating that $J_s^a(x)$ form an SU(2) level $k=2$ Kac-Moody algebra. 
Meanwhile, the charge and flavor currents satisfy 
\begin{eqnarray}
&& [J_c(x),J_c(x')]=2ki\delta'(x-x'), \nonumber \\
&&  [J_f^a(x), J_f^b(x')]
  = i\epsilon^{abc}J_f^a(x)\delta(x-x')
 \nonumber \\
  && \hspace{3cm} +\frac{ki}{4\pi}\delta_{a,b}\delta'(x-x').
\end{eqnarray}
They form a U(1) Kac-Moody and another SU(2) level-2 Kac-Moody algebra, 
separately.

It is now quite natural to introduce a Majorana representation of the
spin current operators in the form,
\begin{eqnarray}
 && J_s^x(x)=-i:\chi_2(x)\chi_3(x):, \nonumber \\
 && J_s^y(x)=-i:\chi_3(x)\chi_1(x):, \nonumber \\
 && J_s^z(x)=-i:\chi_1(x)\chi_2(x):,
\end{eqnarray}
where $\chi_1(x),\chi_2(x),$ and $\chi_3(x)$ are left-moving free Majorana
fermion fields, and it can be shown to reproduce the SU(2) level-2 Kac-Moody
commutation relations.
It is important to note that this representation is only appropriate
for the two channel model as it leads to a level-2 algebra. It would be
{\it inappropriate} for the single channel Kondo model where the corresponding
spin current generates a level-1 algebra.

In a similar way, we can also introduce Majorana representations for the
flavor currents
\begin{eqnarray}
 && J_f^x(x)=-i:\chi'_2(x)\chi'_3(x):, \nonumber \\
 && J_f^y(x)=-i:\chi'_3(x)\chi'_1(x):, \nonumber \\
 && J_f^z(x)=-i:\chi'_1(x)\chi'_2(x):,
\end{eqnarray}
which reproduces the commutation relations satisfied by the flavor currents,
and
\begin{equation}
  J_c(x)=-2i:\chi'_4(x)\chi'_5(x):
\end{equation}
can represent the charge current operator. Note that $\chi'_{\alpha}$ with
$\alpha=1,2,3,4,5$ are also left-moving free Majorana fermion fields.
It is well-known that the dynamics of charge, flavor, and spin are 
completely determined by the commutation relations of 
the current operators. Though the spin currents of the two channel Kondo model
can be represented in terms of three Majorana fermion fields
$\chi_{\alpha}(x)$ ($\alpha=1,2,3$), we emphasize that
they can not be given any simple physical interpretation in terms of the 
original conduction electrons $\psi_{j,\sigma}(x)$.

At this point we have the current operator terms in the Hamiltonian
as quartic in the Majorana fields. The Sugawara construction enables one to
write kinetic energy terms, which are quadratic in field operators, as quartic
terms. This is what was done earlier in writing the free part of the
Hamiltonian in form of Eq.(\ref{suga}), and it is convenient if one is 
pursuing a purely
algebraic approach as used in the conformal field theory \cite{al}. However
for our purposes it is more convenient now to perform an inverse Sugawara
construction by the usual point-splitting procedure again, and rewrite the
terms quartic in the Majorana fermions as  kinetic energy terms which are
quadratic \cite{kz,ginsparg}:
\begin{eqnarray}
&& :J_c(x)J_c(x):
   =4\sum_{\alpha=4}^{5}:\chi'_{\alpha}(i\partial_x)\chi'_{\alpha}(x);
\nonumber \\
&& :\vec{J}_f(x)\cdot\vec{J}_f(x):
   =2\sum_{\alpha=1}^{3}:\chi'_{\alpha}(i\partial_x)\chi'_{\alpha}(x);
\nonumber \\
&& :\vec{J}_s(x)\cdot\vec{J}_s(x):
   =2\sum_{\alpha=1}^{3}:\chi_{\alpha}(i\partial_x)\chi_{\alpha}(x).
\end{eqnarray}
The model Hamiltonian is transformed and divided into the following two parts,
\begin{eqnarray}
&& H_c+H_f=\frac{v_f}{4\pi}\sum_{\alpha=1}^{5}
  \int_{-\infty}^{+\infty}dx :\chi'_{\alpha}(x)(i\partial_x)\chi'_{\alpha}(x):,
\nonumber \\
&& H_s=\frac{v_f}{4\pi}\sum_{\alpha=1}^{3}
    \int_{-\infty}^{+\infty}dx :\chi_{\alpha}(x)(i\partial_x)\chi_{\alpha}(x):
\nonumber \\ && \hspace{3cm}        
      -\frac{iJ}{2}\vec{S}_d \cdot :\vec{\chi}(0)\times\vec{\chi}(0):.
\end{eqnarray}
$H_c+H_f$ describes the non-interacting charge and flavor degrees of freedom.
It has a symmetry group $ U(1)\otimes SU(2)_2 = SO(5) $ and is expressed by
five free Majorana fermion fields $\chi'_{\alpha}(x)$
($\alpha=1,2,3,4,5$). $H_s$ is the main part of the model and describes
the spin degrees of freedom with three left-moving Majorana 
fermion fields $\chi_{\alpha}$ ($\alpha=1,2,3$) interacting with the impurity 
spin. It has the symmetry $SU(2)_2$ or $SO(3)$ 
so that the full Hamiltonian has the symmetry group
$SO(5)\otimes SO(3)= SO(8)$, which is represented by eight different Majorana 
fermion fields.  

In the two channel model Hamiltonian, $H_s$ given in Eq.(11) is the only part 
which includes an interaction with the impurity spin. This part of the 
Hamiltonian is exactly 
equivalent to the vector part of the $\sigma$-$\tau$ model.
The $\sigma$-$\tau$ (compactified) model is defined by
\begin{eqnarray}
&& H'=-t\sum_{n=0}^{\infty}\sum_{\sigma}\left[c_{\sigma}^{\dag}(n+1)
c_{\sigma}(n) +H.c.\right] \nonumber \\ && \hspace{1cm}
+ J \left[\vec{s}(0)+\vec{\tau}(0)\right]\cdot\vec{S}_d,
\end{eqnarray} 
where $\vec{s}(0)$ and $\vec{\tau}(0)$ denote the conduction electron spin and
isospin current operators at the impurity site. The spin currents are  
defined as usual, and form a $SU(2)$ level-1 Kac-Moody algebra. 
The conduction electron isospin currents are defined as follows
\begin{eqnarray}
 && \tau^{+}(n)=(-1)^n c_{\uparrow}^{\dag}(n)c_{\downarrow}^{\dag}(n),
    \hspace{.1cm}
 \tau^{-}(n)=(-1)^n c_{\downarrow}(n)c_{\uparrow}(n), \nonumber \\
 && \tau^{z}(n)=\frac{1}{2}\left[ c_{\uparrow}^{\dag}(n)c_{\uparrow}(n)
        +c_{\downarrow}^{\dag}(n)c_{\downarrow}(n)-1 \right],
\end{eqnarray}
which also forms a level-1 $SU(2)$ Kac-Moody algebra.
Under the particle-hole transformation:
$c_{\uparrow}(n)\rightarrow c_{\uparrow}(n)$ and
$ c_{\downarrow}(n)\rightarrow (-1)^n c_{\downarrow}^{\dag}(n)$,
the conduction electron spin current operators change into the
isospin current operators, and vice versa.
It is known that Majorana fermions can be introduced as follows,
\begin{equation}
 \left( \matrix{ c_{\uparrow}(n)\cr
       c_{\downarrow}(n)} \right)
     =\frac{e^{i\pi n/2}}{\sqrt{2}}
   \left(\matrix{ \Psi_1(n)-i\Psi_2(n)\cr
                 -\Psi_3(n)-i\Psi_0(n)}\right), 
\end{equation}
where a phase factor has been introduced to absorb the staggered phase factors
of the conduction electron states, $\Psi_0(n)$ is referred to as the scalar
component, and $\Psi_1(n)$, $\Psi_2(n)$, $\Psi_3(n)$ are referred to as the  
vector components. The model is thus divided into two parts
\begin{eqnarray}
&& H'=H'_{\rm sc}+H'_{\rm vec} \nonumber \\
&& H'_{\rm sc}=\frac{v_f}{2\pi}
  \int_{-\infty}^{+\infty}dx :\Psi_0(x)(i\partial_x)\Psi_0(x):
\nonumber \\
&& H'_{\rm vec}=\frac{v_f}{2\pi}\sum_{\alpha=1}^{3}
    \int_{-\infty}^{+\infty}dx :\Psi_{\alpha}(x)(i\partial_x)\Psi_{\alpha}(x):
\nonumber \\ && \hspace{3cm}
      -{iJ}\vec{S}_d \cdot :\vec{\Psi}(0)\times\vec{\Psi}(0):.
\end{eqnarray}
We can identify $H'_{\rm vec}$ with $H_s$ up to an overall factor 
$2$. This implies that the application of 
the $\sigma$-$\tau$ model is not restricted to the very low energy regime but 
 can be used to calculate the impurity contribution to the thermodynamics
of the two channel Kondo model over the full temperature range. This result is
{\it exact} subject only to the requirement of linear dispersion for the 
conduction electrons. 


We briefly show that the Majorana fermion form of the Hamiltonian can be
used to confirm the analysis of the many-body excitations
at the low energy fixed point given in earlier work \cite{ml,bhz,ye}.
In our previous paper \cite{bhz} on the isotropic
$\sigma$-$\tau$ and O(3) Anderson models we found that the excitations
at the fixed point could be constructed from free Majorana
fermion excitations from the vector and scalar channels. The boundary
conditions of the vector part had to be changed relative to the scalar
part so that if the scalar part has periodic boundary conditions the
vector part has antiperiodic ones (sector A), and vice versa (sector B). 
This difference
can be interpreted as due to strong coupling with the impurity spin in
the vector channel, but we also found the same fixed point for the compactified
Anderson model for all values of $U$, so it can be interpreted as 
the $U=0$ fixed point of O(3) Anderson model. The combined many-body
spectrum agreed both with the results of the numerical renormalization
group and the conformal field theory: sector A excitations for a chain
with odd number of sites and sector B for a chain with an even number of sites.
These two distinguishable excitation sectors give the same 
low energy physics \cite{bh,bhz}.

For a similar analysis
of the two channel model in the Majorana fermion representation given in Eq.
(11), we have to replace the scalar Majorana fermion in $H_{sc}$ by the 
five free Majorana fermions describing the flavor and charge degrees of 
freedom in $H_f+H_c$. We can then construct the excitation spectrum at
the fixed point with different boundary conditions for the fermions
 in the spin and charge-flavor sectors. To generate
the spectrum as found in the numerical renormalization group calculations
for the two channel case we have to include excitations corresponding to
both sectors A and B.  We note
that the boundary conditions we impose on the fields $\chi_{\alpha}(x)$
cannot be placed in direct correspondence with boundary conditions imposed
on the physical conduction electron fields $\psi_{j,\sigma}(x)$.
The resulting spectrum is
given in Table I, which is in agreement with the numerical renormalization
group and conformal field theory calculations of the two channel Kondo model.
It is the same as that conjectured in our earlier work
\cite{bhz} and in agreement
with the work of Maldacena and Ludwig \cite{ml} and Ye \cite{ye}. 

Away from the low energy fixed point, using the Majorana fermion 
representation, it is straightforward to identify the leading boundary 
operator as we have shown in our previous paper \cite{bhz} for the 
$\sigma$-$\tau$ model.  
Since the interaction is restricted to the spin part of the Hamiltonian with 
an $O(3)$ symmetry group, the boundary operator must be $O(3)$ invariant. 
There is a unique $O(3)$ invariant operator with a scaling
dimension $\frac{3}{2}$ in the spin part: $\chi_1(x)\chi_2(x)\chi_3(x)$. 
Therefore, we are led to identify $\theta\chi_1(0)\chi_2(0)\chi_3(0)$ as 
the leading boundary operator of the two channel Kondo model \cite{coleman},
where $\theta$ is an anticommuting variable. As its scaling dimension 
is greater than one, this boundary operator is irrelevant. The 
impurity contributions to the thermodynamic properties  
of the two channel Kondo model can be obtained by the second order 
perturbation calculations of this leading irrelevant boundary operator.

To summarize, we have used a Majorana fermion representation of the spin 
currents to obtain an exact equivalent Majorana fermion description of the
two channel Kondo model. The spin part corresponds to the vector part of the 
$\sigma$-$\tau$ model. It provides a simple prescription for understanding
the low-energy fixed point. This formulation makes possible further 
developments such as the calculation of spin-spin current dynamical 
correlations for the two channel model from the $\sigma$-$\tau$ model 
alone.

\bigskip
\noindent{\bf Acknowledgment}\par
\bigskip
We thank Dr. Y. Chen for helpful suggestions,
and we are also grateful to the EPSRC for the support of a research grant
(No. GR/J85349).

\begin{table}
\caption{Many-body excitation energies and their
  corresponding degeneracies ($dg$) of the two channel Kondo model
corresponding to the single particle spectrum at the low-energy 
fixed point. The energy levels without primes correspond to sector A, 
and those with primes to sector B. } 

\begin{tabular}[t]{cccc}
   $E_{\rm ex}/(\pi v_{\rm F}/l)$  & $\sum_k n_k \epsilon_k $
       & $dg$   & total $dg$\\
  \hline
   0 & 0 & 2 & 2 \\
  \hline
   1/8 & $0'$ & 4 & 4 \\
   \hline
   1/2 &$\varepsilon_{1/2}$ & 10 & 10 \\
  \hline
  5/8  &$\varepsilon'_{1/2}$ & 12 & 12 \\
  \hline
   1 &$\varepsilon_{1}$ & 6 &  \\
     &$2\varepsilon_{1/2}$ & 20 & 26 \\
  \hline
  9/8 &$\varepsilon'_{1}$ & 20 &  \\
     &$2\varepsilon'_{1/2}$ & 12 & 32 \\
  \hline
   3/2 &$\varepsilon_{3/2}$ & 10 &  \\
       &$\varepsilon_{1} + \varepsilon_{1/2}$ & 30 & 60 \\
       &$3\varepsilon_{1/2}$ & 20 &  \\
   \hline
   13/8 &$\varepsilon'_{3/2}$ & 12 &  \\
       &$\varepsilon'_{1} + \varepsilon'_{1/2}$ & 60 & 76 \\
\end{tabular}
\end{table}


\begin{thebibliography}{99}
\bibitem{cz} D. L Cox and A. Zawadowski, to be published in Rev. Mod. Phys. 
  (1997); cond-mat/9704103.

\bibitem{andrei}N. Andrei, and C. Destri, Phys. Rev. Lett. {\bf 52}, 364
(1984); A. M. Tsvelik and P. B. Wiegmann, J. Stat. Phys. {\bf 38}, 125
(1984).

\bibitem{pc}D. M. Cragg, P. Lloyd, Ph. Nozi\`eres,
  J. Phys. C, {\bf 13}, 803 (1980);
H. B. Pang, D. L. Cox, Phys. Rev. B, {\bf 44}, 9454 (1991).

\bibitem{al}I. Affleck and A. W. W. Ludwig, Nucl. Phys. B {\bf 360}, 641
(1991): Phys. Rev. B {\bf 48}, 7297 (1993);
I. Affleck, A. W. W. Ludwig, H. -B. Pang, D. L. Cox,
   Phys. Rev. B, {\bf 45}, 7918 (1992).

\bibitem{ek}V. J. Emery and S. Kivelson, Phys. Rev. B {\bf 46}, 10812
(1992).

\bibitem{ml} J. M. Maldacena and A. W. W. Ludwig, to be published in 
  Nucl. Phys. B, (1997); cond-mat/9502109.

\bibitem{cit} P. Coleman, L. Ioffe, and A. M. Tsvelik, Phys. Rev. B {\bf 52},
6611 (1995).

\bibitem{cs}P. Coleman and A. J. Schofield, Phys. Rev. Lett. {\bf 75},
 2184 (1995).

\bibitem{zh}G. -M. Zhang and A. C. Hewson, Phys. Rev. Lett. {\bf 76},
 2137 (1996); Phys. Rev. B {\bf 54}, 1169 (1996).

\bibitem{bh} R. Bulla and A. C. Hewson,
    to be published in  Z. Phys. {\bf B} (1997), cond-mat/9701152.

\bibitem{bhz}R. Bulla,  A. C. Hewson, and G. -M. Zhang, submitted to Phys. Rev.
 B, (1997), cond-mat/9704024.

\bibitem{kz}V. G. Knizhnik and A. B. Zamolodchikov, Nucl. Phys. B {\bf 247},
83 (1984).

\bibitem{ginsparg} P. Ginsparg in {\it Fields, Strings and Critical Phenomena},
Les Houches XLIX, E. Br\`ezin and J. Zinn-Justin (Eds), (North
Holland, Amsterdam 1990).

\bibitem{ye}J. Ye, preprint, cond/mat9609057. 

\bibitem{coleman} A similar leading irrelevant boundary operator was derived 
  for the compactified two channel Kondo model, see ref.7.
  
\end{thebibliography}
\end{document}